\documentclass[10pt,conference]{IEEEtran}
\IEEEoverridecommandlockouts

\usepackage{cite}
\usepackage{caption}
\usepackage{placeins} 
\usepackage{cuted}

\usepackage{float}
\usepackage{amsmath,amssymb,amsfonts}
\usepackage{algorithm}
\usepackage{algorithmicx}
\usepackage{algpseudocode}
\usepackage{graphicx}
\usepackage{textcomp}
\usepackage{xcolor}
\def\BibTeX{{\rm B\kern-.05em{\sc i\kern-.025em b}\kern-.08em
    T\kern-.1667em\lower.7ex\hbox{E}\kern-.125emX}}

\title{Hybrid NOMA Assisted Heterogeneous Semantic and Bit Users Communication}
\author{\IEEEauthorblockN{Ishtiaque Ahmed, and Leila Musavian}
\thanks{The authors are with the School of Computer Science and Electronic Engineering, University of Essex, Wivenhoe Park, Colchester CO4 3SQ, United Kingdom (e-mail: { \{ishtiaque.ahmed, leila.musavian\}@essex.ac.uk). This work has been submitted to the IEEE for possible publication. Copyright may be transferred without notice, after which this version may no longer be accessible.} }
 }
    
\begin{document}

\maketitle

\begin{abstract}
In this paper, we utilize a downlink hybrid Non-Orthogonal Multiple Access (NOMA) framework to support multiple semantic and bit users within the communication network. The hybrid NOMA setup exploits both NOMA and Orthogonal Multiple Access (OMA) which has the benefit of enhancing Spectral Efficiency (SE) by allowing users to dynamically access the resources in multiple heterogeneous slots. This enables integrating semantic and bit users based on their channel gains, while adopting bit-to-semantic decoding order in slots including heterogeneous users. An optimization problem for the power allocation is formulated with the aim of maximizing the equivalent ergodic semantic SE with a constraint on the total available power of the Access Point (AP). The proposed algorithm uses NOMA in shared slots and OMA in bit-user-only slots. Simulation results validate the benefits of heterogeneous users hybrid NOMA setup in comparison to OMA-only for heterogeneous users.
\end{abstract}

\section{Introduction} \label{intro}
With the exponential surge in wireless data traffic, enabling massive connectivity while ensuring the required transmission capacity poses a significant challenge for the development of next-generation communication networks \cite{wang2024adaptive}. On the other hand, building upon the foundational work by Shannon and Weaver, semantic communication emerges as a novel paradigm that revolutionizes communication, focusing on transmitting the semantic content of information, and is particularly suited in low Signal-to-Noise Ratio (SNR) environments \cite{yang2022semantic}. However, at higher SNR regimes, traditional bit-based communication can achieve near-lossless data recovery, making semantic communication less beneficial. This suggests that semantic communication should complement and coexist with the traditional Shannon communication to enhance the overall performance \cite{yan2022resource}. This coexistence of semantic and bit-based communications within the same network defines heterogeneity within the users set in the network. In fact, with the evolution of semantic communication, new challenges emerge in systems design, particularly in implementing Successive Interference Cancellation (SIC) for supporting heterogeneous users within the network \cite{chen2023uplink}.

To support massive connectivity, next-generation communication systems necessitate advanced multiple access techniques that push the boundaries of classical schemes such as Orthogonal Multiple Access \cite{wang2020thirty}. In this respect, Non-Orthogonal Multiple Access (NOMA) is considered a promising technology for enhancing the Spectral Efficiency (SE) of wireless systems by allowing multiple users to share the same time-frequency resources through power or code-domain multiplexing. However, NOMA also presents challenges which need to be carefully addressed, such as increased interference management complexity and fairness concerns \cite{pei2022next,you2021towards}. The increase in achievable SE using NOMA depends on the specific system configuration and use case, as NOMA does not always outperform orthogonal schemes, particularly in heterogeneous users communications \cite{mu2022heterogeneous}. In traditional NOMA, SIC is performed based on channel conditions, where the stronger user decodes the signal of the weaker user to mitigate interference. However, this approach cannot be directly applied in joint bit and semantic users communication systems. This is primarily due to the requirement of pre-trained neural networks at the semantic encoder and decoder for successful transmission and reception \cite{xie2021deep}. Consequently, the bit user is unable to decode the semantic signal from the superimposed signal, as traditional bit-based communication does not require any prior training at the transmitter or receiver. In contrast, the semantic user can decode the bit-based signal, resulting in a bit-to-semantic decoding order for joint bit and semantic users communication \cite{mu2022semantic}.

Although conventional NOMA enables multiple access through SIC, its decoding strategy must be adapted to support joint bit and semantic users. Given the differences in their signal processing requirements, a more flexible approach is needed to optimize users pairing and interference management. In this context, hybrid NOMA can exploit the advantages of both OMA and pure NOMA through seamless transitions from one scheme to the other \cite{ding2024next}. The crux of hybrid NOMA is to enable multiple users occupy and share the same resource via multiple time slots, for enhancing the overall system capacity. In addition to addressing SIC challenges, optimal power allocation plays a crucial role in the hybrid NOMA framework. By efficiently distributing power among bit and semantic users, the total expected semantic rate can be maximized while ensuring the bit-based users minimum rate requirements, as in \cite{mu2022semantic}.

To fully leverage the benefits of semantic communication within hybrid NOMA, it is important to understand how the semantic communication is characterized. The rate at which semantic information is transmitted with a prescribed accuracy is referred to as the semantic rate, which depends on the semantic similarity function with value ranging between zero and one \cite{yan2022resource}. This function quantifies the resemblance between original signal at the transmitter and reconstructed one at the receiver. When semantic similarity function tends to one, the signal will increasingly resemble the original bit-based signal rather than its reconstructed semantic counterpart \cite{xie2021deep}.

Since the semantic similarity function plays a crucial role in determining the effectiveness of semantic communication, a learning-based framework is needed to extract and reconstruct semantic features. To this end, a deep learning based system known as DeepSC is developed \cite{xie2021deep} which consists of semantic encoder/decoder and channel encoder/decoder. The DeepSC is equipped with neural networks for semantic features extraction. Semantic similarity function depends on neural network structure in DeepSC and the channel condition, and is based on the match level between the transmitted and received sentences using the Biderctional Encoder Representations from Transformers (BERT) model \cite{xie2021deep}. The pre-trained BERT model \cite{devlin2018bert} captures the context from preceding and succeeding words in a sentence.

The works in \cite{mu2022semantic} and \cite{ding2022hybrid} have advanced the integration of semantic and bit-based communications and the deployment of hybrid NOMA. In \cite{mu2022semantic}, a semi-NOMA framework is proposed that investigates the trade-offs between semantic and bit-based communications in a shared transmission scheme. However, this approach does not exploit hybrid NOMA with user pairing based on channel conditions, limiting its adaptability in dynamic wireless environments. In \cite{ding2022hybrid}, a multi-user hybrid NOMA scheme is introduced that optimizes power allocation and minimizes energy consumption under latency constraints. However, this work focuses solely on bit-based communication and does not incorporate semantic communication.

Inspired by the above works, we harness the benefits of heterogeneous users by adopting hybrid NOMA and a tailored decoding strategy that improves interference management. The key contributions of this work are: i) we propose a hybrid NOMA framework for heterogeneous users that dynamically assigns bit and semantic users to time slots based on their channel conditions. To efficiently implement this, we develop a user pairing and decoding strategy, outlined in Algorithm 1 which enhances SE while ensuring adaptive resource allocation; ii) unlike previous works that assume a fixed decoding order, we deploy an adaptive decoding strategy that optimally manages interference in hybrid NOMA. In particular, our approach ensures a bit-to-semantic decoding order in time slots where both bit and semantic users coexist, while applying stronger-to-weaker SIC decoding in time slots containing only semantic users; iii) we formulate and solve a power allocation optimization problem that maximizes the equivalent ergodic semantic SE with a constraint on the total available power; iv) we demonstrate through simulations that the optimization prioritizes bit user in a heterogeneous users network by allocating more power, particularly at higher available power values.

\section{System Model}
\begin{figure}[t]
\centering
\includegraphics[trim={5cm 2.05cm 4.05cm 7cm},clip,width=1.001\columnwidth]{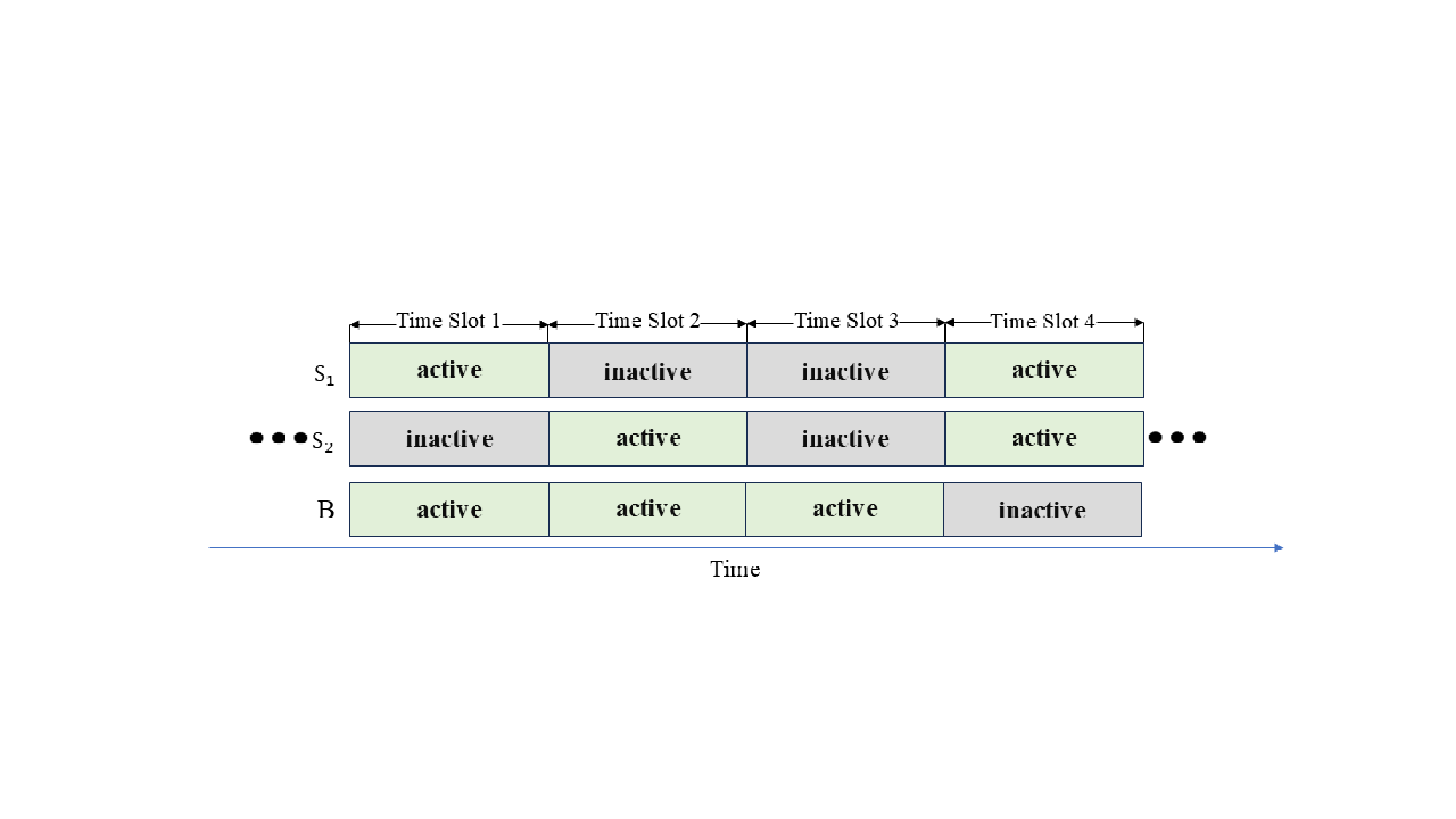}
\vspace{-50pt}
\caption{Users allocation in hybrid NOMA.}
\label{fig1}
\end{figure}

We consider a downlink hybrid NOMA communication system consisting of a bit user (B) and two semantic users ($\text{S}_1$ and $\text{S}_2$), sharing the entire time-frequency resources. In our proposed approach, we allow at most two users to share any slot, as shown in Fig.~\ref{fig1}. Communication between the AP and the users occurs over Rayleigh fading channels, where the channel gains remain constant within each time slot but may vary across time slots. At the receiver, additive white Gaussian noise (AWGN) with mean zero and variance $\sigma^2$ is assumed to impair the superimposed signal. We deploy the DeepSC model in our hybrid NOMA assisted system that can be generally applied to any semantic architecture.

Let $P_{\max}$ be the total available power budget at the AP. The channel gain coefficients from the AP to users B, $\text{S}_1$, and $\text{S}_2$ are denoted by $h_{\text{B}}$, $h_{\text{S1}}$, and $h_{\text{S2}}$, respectively, while the transmit powers allocated to them are $P_{\text{B}}$, $P_{\text{S1}}$, and $P_{\text{S2}}$. The semantic rate for any semantic user is given as \cite{yan2022resource}
\begin{equation}
R_{\text{S}i}(K, L) = \frac{WI}{KL} \epsilon(K, \gamma), i \in \{1, 2\}
\label{semanticrate}
\end{equation}
where $\epsilon (K, \gamma)$ is the semantic similarity function, $W$ is the channel bandwidth, $I$ is the amount of semantic information in any transmitted message in units of semantic (suts), $K$ represents the average number of semantic symbols transmitted for each word through DeepSC \cite{xie2021deep}, $\gamma$ represents the received SNR, and $L$ denotes the number of words per sentence. With varying values of $K$ and $\gamma$, $\epsilon (K, \gamma)$ was found to be monotonically non-decreasing with $\gamma$ \cite{yan2022resource}. Moreover, its gradient change increases first with $\gamma$ and then decreases, suggesting a sigmoid shape pattern for $\epsilon (K, \gamma)$. Authors in \cite{mu2022heterogeneous} deployed the data-regression method to tractably approximate the values of $\epsilon (K, \gamma)$ with a generalized logistic function (common form of sigmoid function), as
\begin{equation}
\epsilon (K, \gamma) \overset{\triangle}{=} A_{K,1} + 
\frac{A_{K,2} - A_{K,1}}{1 + e^{-(C_{K,1} \gamma + C_{K,2})}},
\label{logisticapprox}
\end{equation}
where the lower (left) asymptote, upper (right) asymptote, growth rate, and the mid-point parameters of the logistic function are respectively denoted by $A_{K,1}$, $A_{K,2}$, $C_{K,1}$, and $C_{K,2}$ for different values of $K$. These parameters are determined by following the minimum mean square criterion for fitting the logistic function to $\epsilon (K, \gamma)$. To ensure that semantic communication achieves desired accuracy at the receiver, $\epsilon (K, \gamma)$ must meet a minimum threshold $\epsilon_{\mathrm{th}}$, below which the transmission will be treated as invalid. This is represented as
\begin{equation}
\epsilon_K(\gamma) = 
\begin{cases} 
\epsilon(K, \gamma), & \text{if } \epsilon(K, \gamma) \geq \epsilon_{\mathrm{th}}, \\[10pt]
0, & \text{otherwise}.
\end{cases}
\label{logisticapprox2}
\end{equation}
The approximated generalized logistic function depends explicitly on $\gamma$, while the effect of $K$ is captured through the parameters $A_{K,1}$, $A_{K,2}$, $C_{K,1}$, and $C_{K,2}$.

We denote the rates of users $\text{S}_1$, $\text{S}_2$ and B by $R_{\text{S1}}$, $R_{\text{S2}}$, and $R_{\text{B}}$, respectively, where $R_{\text{S1}}$ and $R_{\text{S2}}$ are in suts/s/Hz, while $R_{\text{B}}$ is in bits/s/Hz. The rate of User B is determined on the basis of Shannon's capacity formula. Perfect channel state information is assumed to be available at the AP. To ensure efficient decoding of signals, the AP informs the receiver about the type of active users in each time slot via embedded control information in the headers of transmitted data packets.

Our goal is to maximize the expected equivalent semantic SE of the three users while ensuring that the system obeys the AP power constraint. We transform the rate of bit user $R_{\text{B}}$ into its semantic counterpart $R_{\text{SB}}$ in suts/s/Hz, as in \cite{yan2022resource}, yielding
\begin{equation}
R_{\text{SB}}=R_{\text{B}} \frac{I}{\mu L} \epsilon_C,
\label{equivalentsemanticrate}
\end{equation}
where $\mu$ represents the average number of bits per word, the semantic similarity function for User B is denoted by $\epsilon_C$ and taken as one to reflect error-free transmission. This transformation allows a meaningful comparison between both types of users by expressing their efficiencies in the same unit.

For the hybrid NOMA structure, in each time slot, up to two users are scheduled to be active and share the resources. Users' activity is governed by the indicator variables $I_{\text{B}}$, $I_{\text{S1}}$, and $I_{\text{S2}}$ for users B, $\text{S}_1$, and $\text{S}_2$, respectively. The selection of users for pairing is done dynamically based on real-time channel conditions, ensuring that the strongest user pairs with the weakest one across different time slots without a fixed ordering. When a semantic user has stronger channel gain than the bit user, the slot is allocated exclusively to the bit user, effectively operating as OMA. In contrast, when the channel conditions allow simultaneous transmission, NOMA is employed, where both users share the same time-frequency resources while following a bit-to-semantic decoding order. This adaptive hybrid NOMA strategy dynamically transitions between OMA and NOMA while maintaining the bit-to-semantic decoding strategy.

As semantic and bit users can simultaneously be active in each slot, a new decoding mechanism will be required. Let us denote the corresponding transmit signals for users B, $\text{S}_1$, and $\text{S}_2$ by $x_{\text{B}}$, $x_{\text{S1}}$, and $x_{\text{S2}}$, respectively. In slots where any two users share the channel via NOMA, the received signal will be represented as a superposition of their transmitted signals, scaled by their respective channel gains and corrupted by AWGN. In such slots, decoded signal of one user will always be affected by interference from the other. For the OMA slots, there will be no decoding interference experienced. Moreover, traditional channel-condition based SIC will no longer be beneficial in joint bit and semantic users communication networks as the bit user cannot decode a semantic signal due to the absence of prior training at transmitter and receiver. Consequently, we adopt the bit-to-semantic decoding order in slots involving both types of users, where semantic signals are decoded interference-free after SIC. However, the bit user being decoded first will treat the semantic user's signal as interference.

The pairing approach and decoding strategy for users are outlined in Algorithm 1.
\begin{algorithm}[]
\caption{User Pairing and Decoding Strategy for the Hybrid NOMA Framework}
\begin{algorithmic}[1] 
    \State \textbf{Initialization:} Set $P_{\max}$ value, and $I_{\text{B}}=0$, $I_{\text{S1}}=0$, $I_{\text{S2}}=0$.
    \State \textbf{Users Pairing in Each Slot:} Determine $h_{\text{B}}$,  $h_{\text{S1}}$,  $h_{\text{S2}}$ and pair the strongest user with the weakest one. Set the indicators for the paired users to 1.
    \If{slot contains B and either $\text{S}_1$ or $\text{S}_2$}
     \If{involved semantic user has a stronger channel power gain than User B}
            \State Reset its indicator to 0 which allocates the slot to User B only (OMA). 
            \Else \State follow \textbf{Bit-to-semantic Decoding:}
            \State \textbf{B-receiver} decodes its own signal treating semantic user's signal as noise.
            \State \textbf{S-receiver} firstly decodes B user's signal and applies SIC to subtract it, resulting in its own signal without interference.
            \EndIf
    \ElsIf{slot contains $\text{S}_1$ and $\text{S}_2$}
            \State \textbf{Stronger Semantic User} decodes its own signal treating the weaker user's signal as noise.
            \State \textbf{Weaker Semantic User} firstly decodes the stronger user's signal and applies SIC to subtract it, resulting in its own signal without interference.
        \EndIf
\end{algorithmic}
\end{algorithm}

\section{Optimal Power Allocation and Users Detection}
After pairing users and establishing the decoding order in each time slot, we now formulate the power allocation optimization problem.
\subsection{Power Allocation Optimization}
Objective of our optimization problem is to maximize the expected value of the equivalent semantic SE, which can be formulated as

\begin{equation}
\begin{aligned}
\max _{\{P_{\text{B}}, P_{\text{S1}}, P_{\text{S2}}\}}  \mathbb{E}\left[I_{\text{S1}} R_{\text{S1}}+I_{\text{S2}} R_{\text{S2}}+I_{\text{B}} R_{\text{SB}}\right],
\end{aligned}
\label{convexobjfunc}
\end{equation}
\begin{equation}
\begin{aligned}
\text { s.t. } \quad & P_{\text{B}} + P_{\text{S1}} + P_{\text{S2}} \leq P_{\max}, \\
& P_{\text{B}},\; P_{\text{S1}},\; P_{\text{S2}} \geq 0
\end{aligned}
\label{convexpowcons}
\end{equation}
where $\mathbb{E} [.]$ represents the expectation operator. Each of the considered rates in \eqref{convexobjfunc} can be generalized with a closed-form approximation of \eqref{logisticapprox}, exhibiting a logistic distribution model due to semantic similarity function \cite{mu2022heterogeneous}, where $A_{K,1}=0.37$, $A_{K,2}=0.98$, $C_{K,1}=0.25$, and $C_{K,2}=-0.7895$.

Unlike conventional rate functions, the approximated semantic similarity function does not exhibit concavity with respect to $\gamma$ which renders the optimization problem as non-convex. However, the problem satisfies the ``time-sharing" condition following the proof in \cite{mu2023exploiting} and \cite{yu2006dual}. If any problem satisfies the ``time-sharing" condition \cite{yu2006dual}, then strong duality holds between the primal and Lagrangian dual problems, regardless of convexity of the original problem. Moreover, when strong duality exists, the Karush-Kuhn-Tucker (KKT) conditions are both sufficient and necessary for optimality \cite{boyd2004convex}.

\subsection{Solution Method}
We solve the power allocation problem in \eqref{convexobjfunc} by developing its Lagrangian and then using the KKT conditions in each time slot. The complementary slackness and primal feasibility conditions ensure that the solution respects the power constraint and precludes any negative power allocation.
The KKT conditions are developed based on the Lagrangian formulation given by
\begin{equation}
\begin{aligned}
& \mathcal{L}(P_{\text{B}}, P_{\text{S1}}, P_{\text{S2}}, \lambda)\\
& = \mathbb{E}\left[I_{\text{S1}} R_{\text{S1}}+I_{\text{S2}} R_{\text{S2}}+I_{\text{B}} R_{\text{SB}}\right]+\\
&\lambda\left(P_{\max }-P_{\text{B}}-P_{\text{S1}}-P_{\text{S2}}\right),
\label{lagrange}
\end{aligned}
\end{equation}
where $\lambda$ is the corresponding Lagrange multiplier for the power constraint.

\subsection{Decoding Strategy in Slots with Semantic and Bit Users}
Bit-to-semantic decoding order is followed where the rate of User B will always be adversely impacted by interference from the semantic user. The roots of \eqref{partialPB} provide the solution for optimal power of the bit user.
\begin{equation}
\frac{
\partial \left( \mathbb{E} \left[ I_{\text{B}} R_{\text{SB}} + I_{\text{S}i} R^{\text{SB}}_{\text{S}i} \right] \right)
}{
\partial P_{\text{B}}
}=0, \quad i \in \{1, 2\}
\label{partialPB}
\end{equation}
Specifically, the corresponding rate for bit user in such time slots is defined by
\begin{equation}
R_{\text{B}i}=\log _2\left(1+\frac{P_{\text{B}}\left|h_{\text{B}}\right|^2}{\sigma^2+P_{\text{S}i}\left|h_{\text{B}}\right|^2}\right), i \in \{1, 2\}
\label{RBi}
\end{equation}
where $P_{\text{S}i}$ denotes the power of semantic user paired with User B. Although we consider the equivalent semantic rate of bit user in our objective function, $R_{\text{SB}}$ is merely a scalar multiple of conventional rate $R_{\text{B}i}$. User B first decodes its own signal, while the semantic user extracts its desired signal without interference through SIC. The corresponding rate for semantic user involved in such slots is given by \eqref{RSi}.
\begin{equation}
R^{\text{SB}}_{\text{S}i}=\frac{I}{K L} \epsilon_K\left(\frac{P_{\text{S}i}\left|h_{\text{S}i}\right|^2}{\sigma^2}\right), i \in \{1, 2\}
\label{RSi}
\end{equation}
Say, we solve for $P_{\text{B}}$ when $\text{S}_1$ is active by putting $I_{\text{B}}=1$, $I_{\text{S1}}=1$, and $I_{\text{S2}}=0$. In the considered time slot with B and $\text{S}_1$ active, the closed-form solution for optimum $P_{\text{B}}$ can be found from
\begin{equation}
\begin{aligned}
&\frac{I\left|h_{\text{B}}\right|^2}{\mu L\ln(2)(\sigma^2 + \left|h_{\text{B}}\right|^2P_{\max } - \left|h_{\text{B}}\right|^2P_{\text{B}})}-\\
&\frac{I\left|h_{\text{S1}}\right|^2(A_{K,2}-A_{K,1})C_{K,1}\,e^{(-C_{K,1}\frac{(P_{\max}-P_{\text{B}})\left|h_{\text{S1}}\right|^2}{\sigma^2}+C_{K,2})}}{K L \sigma^2(1+e^{(-C_{K,1}\frac{(P_{\max}-P_{\text{B}})\left|h_{\text{S1}}\right|^2}{\sigma^2}+C_{K,2})})^2}=0
\label{PB}
\end{aligned}
\end{equation}
Similarly, taking the partial derivative with respect to $P_{\text{S1}}$ in such slots and solving for $P_{\text{B}}$ leads to a function with transcendental value \cite{bernard2017formalization} for the optimal power $P^*_{\text{B}}$ of bit user.
\begin{equation}
\begin{aligned}
&f(P_{\text{B}}) = \frac{\left|h_{\text{B}}\right|^2}{\mu \ln(2)(\sigma^2 + \left|h_{\text{B}}\right|^2P_{\max } - \left|h_{\text{B}}\right|^2P_{\text{B}})}-\\
&\frac{\left|h_{\text{S1}}\right|^2(A_{K,2}-A_{K,1})C_{K,1}\,e^{(-C_{K,1}\frac{(P_{\max}-P_{\text{B}})\left|h_{\text{S1}}\right|^2}{\sigma^2}+C_{K,2})}}{K \sigma^2\Bigg(1+e^{(-C_{K,1}\frac{(P_{\max}-P_{\text{B}})\left|h_{\text{S1}}\right|^2}{\sigma^2}+C_{K,2})}\Bigg)^2 }
\end{aligned}
\end{equation}
In such time slots, if the involved semantic user does not satisfy the set minimum threshold $\epsilon_{\mathrm{th}}$, the entire available power is allocated to the bit user. The optimal power allocation for the bit is thus given as
\begin{equation}
P^*_{\text{B}} =
\begin{cases}
f(P_{\text{B}}), & \text{if } \epsilon_K(\gamma) \geq \epsilon_{\mathrm{th}}, \\
P_{\max}, & \text{otherwise}.
\end{cases}
\end{equation}

\subsection{Decoding Strategy in Slots with Semantic Users Only}
In these slots, the Lagrangian is accordingly adjusted as
\begin{equation}
\begin{aligned}
& \mathcal{L'}(P_{\text{S1}}, P_{\text{S2}}, \lambda)\\
& =\mathbb{E}\left[I_{\text{S1}} R_{\text{S1}}+I_{\text{S2}} R_{\text{S2}}\right]+\lambda\left(P_{\max }-P_{\text{S1}}-P_{\text{S2}}\right),
\label{lagrange'}
\end{aligned}
\end{equation}
focusing solely on the rates of semantic users under the power constraint. Performing SIC becomes challenging in such slots due to the joint training of semantic encoders and decoders. To address this issue, we exploit an innovative channel reconstruction-based SIC method proposed in \cite{chen2023uplink}. This method allows the semantic user with better channel condition to be decoded first while treating the signal from the weaker user as noise. The stronger user's signal is reconstructed by reversing the channel effects and mitigating interference until a target semantic similarity $\epsilon_{\mathrm{th}}$ is attained. Once the stronger signal is successfully reconstructed, the weaker semantic user can subtract the decoded stronger user’s signal from the superimposed signal at the receiver. This approach remains effective even when the semantic users have similar channel conditions and rate requirements, allowing either user to be decoded first.

Putting $I_{\text{S1}}=1$, $I_{\text{S2}}=1$, and assuming $\text{S}_2$ experiences higher channel gain than $\text{S}_1$, then the roots of \eqref{partialPS} provide the optimal solution for $P_{\text{S}2}$.
\begin{equation}
\frac{
\partial \left( \mathbb{E} \left[ R_{\text{S}2} + R_{\text{S}1} \right] \right)
}{
\partial P_{\text{S}2}
}=0
\label{partialPS}
\end{equation}
More generally, the achievable semantic rate of the first decoded user in such time slots is given by
\begin{equation}
\begin{aligned}
 R_{\text{S}2} = \frac{I}{K L} \epsilon_K\left( \frac{P_{\text{S}2}\left|h_{\text{S}2}\right|^2}{P_{\text{S}1}\left|h_{\text{S}2}\right|^2 + \sigma^2}\right),
\end{aligned}
\label{RS1slot3}
\end{equation}
Expressing $P_{\text{S}1} = P_{\max }-P_{\text{S}2}$, the closed-form solution for the optimal power of stronger semantic user can be found from
\begin{equation}
\begin{aligned}
\frac{I}{K L} \Bigg[
& \frac{\gamma_2^2 \left( \sigma^2 + P_{\max} \left|h_{\text{S}1}\right|^2 \right)\left(A_{K, 2}-A_{K, 1}\right) C_{K, 1} 
e^{-\left(C_{K, 1} \gamma_2+C_{K, 2}\right)}}
{P_{\text{S}2}^2 \left|h_{\text{S}2}\right|^2\left(1+e^{-\left(C_{K, 1} \gamma_2+C_{K, 2}\right)}\right)^2}\\
& - \frac{\gamma_1 \left(A_{K,2} - A_{K,1}\right) C_{K,1} 
e^{- \left(C_{K,1} \gamma_1 + C_{K,2} \right)}}
{(P_{\max} - P_{\text{S}2}) \left(1 + e^{- \left(C_{K,1} \gamma_1 + C_{K,2} \right)} \right)^2}
\Bigg]=0
\label{strongerPartial}
\end{aligned}
\end{equation}
\vspace*{1.7em}
where $\gamma_1=\frac{\left(P_{\max }-P_{\text{S}2}\right)\left|h_{\text{S}1}\right|^2}{\sigma^2}$, and $\gamma_2=\frac{P_{\text{S}2}\left|h_{\text{S}2}\right|^2}{\left(P_{\max }-P_{\text{S}2}\right)\left|h_{\text{S}2}\right|^2+\sigma^2}$ denote the received SNR at $\text{S}_1$ and $\text{S}_2$, respectively. 
Solving for $P_{\text{S}2}$ after taking the partial derivative with respect to $P_{\text{S1}}$ results in a transcendental value function for the optimal power of stronger semantic user.

\section{Simulation Results}
In this section, we present the simulation figures and analyze the results for the proposed heterogeneous users hybrid NOMA communication setup. Unless specified otherwise, we use the parametric values of $\epsilon_{\mathrm{th}}=0.9$, $K=5$, $\mu=40$, and $I/L=1$. Users are randomly distributed for each Monte Carlo simulation within a 100 meters cell radius. Users channel coefficients for the independently distributed Rayleigh fading are determined based on the distance-dependent path-loss model $\rho=\rho_0(1 / d)^\beta$. We assume a reference path-loss of $\rho_0=-30$ dB at 1 meter, and the path-loss exponent $\beta=4$. All the simulation results are averaged over $10^5$ Monte Carlo channel realizations.

\begin{figure}[!t]
\centering
\includegraphics[trim={0cm 0cm 0cm 0cm},clip,width=1\columnwidth]{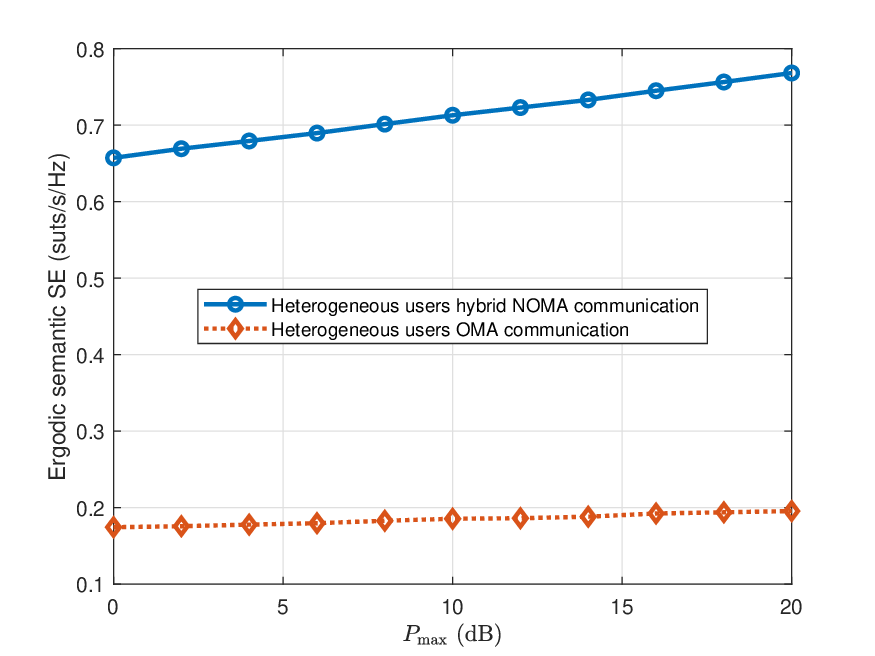}
\caption{Ergodic semantic SE versus $P_{\max}$ with $\epsilon_{\mathrm{th}}=0.9$.}
\label{fig2}
\end{figure}

In the first simulation figure, we compare the performance of the heterogeneous users hybrid NOMA communication with a heterogeneous users OMA setup, where the OMA time fractions are proportionally allocated to each user according to their participation ratio under hybrid NOMA. Fig.~\ref{fig2} shows the equivalent ergodic semantic SE versus $P_{\max}$ for both setups. The results demonstrate that the proposed setup consistently outperforms its OMA counterpart. This indicates the effectiveness of hybrid NOMA for integrating both semantic and traditional bit-based communications, leveraging NOMA and OMA resource allocation strategies. 

\begin{figure}[!t]
\centering
\includegraphics[trim={0cm 0cm 0cm 0cm},clip,width=1\columnwidth]{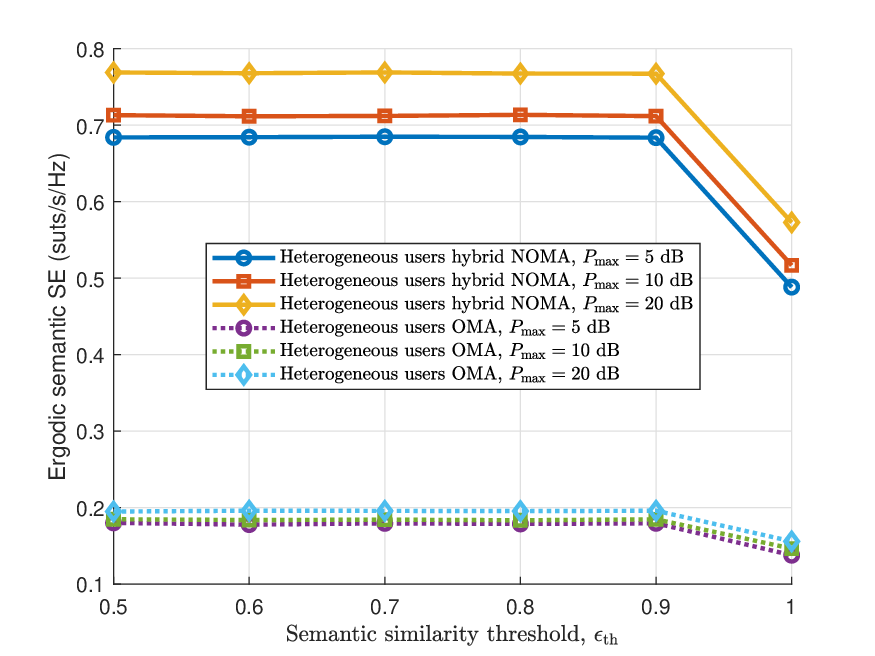}
\caption{Ergodic semantic SE versus $\epsilon_{\mathrm{th}}$ for different $P_{\max}$.}
\label{fig7}
\end{figure}

In Fig.~\ref{fig7}, the equivalent ergodic semantic SE is plotted versus $\epsilon_{\mathrm{th}}$ for $P_{\max}$ values of 5 dB, 10 dB, and 20 dB. It is evident that the proposed setup achieves higher ergodic semantic SE across all values of $\epsilon_{\mathrm{th}}$ for different $P_{\max}$. For both the setups as the value of $\epsilon_{\mathrm{th}}$ approaches one, a drop in SE is recorded due to the inherent limitations in accurate semantic reconstruction at higher semantic similarity thresholds. However, the hybrid NOMA still maintains a higher SE due to the higher resource utilization by the bit user aiding in accurate signal reconstruction. 

\begin{figure}[!t]
\centering
\includegraphics[trim={0cm 0cm 0cm 0cm},clip,width=1\columnwidth]{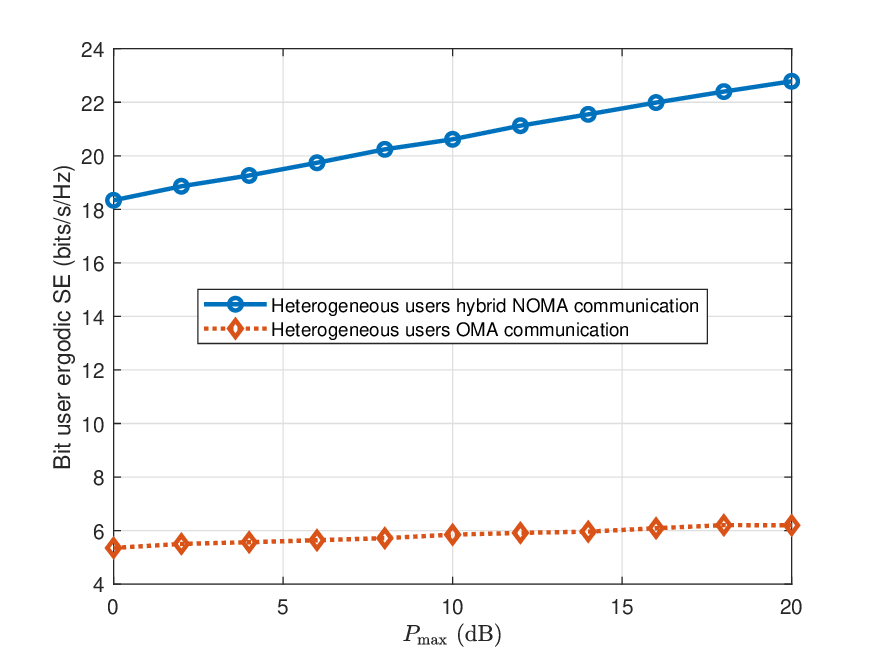}
\caption{Bit user ergodic SE versus $P_{\max}$ with $\epsilon_{\mathrm{th}}=0.9$.}
\label{figBSE}
\end{figure}

Fig.~\ref{figBSE} shows the bit user ergodic SE versus $P_{\max}$ for hybrid NOMA and OMA setups. As $P_{\max}$ increases, the hybrid NOMA curve consistently lies above the OMA curve, demonstrating the benefits of its flexible resource allocation strategy. If the semantic user has stronger channel gain than the bit user in shared time slots, the entire power is redirected to the bit user, otherwise the total power splits between both the users. 

\begin{figure}[!t]
\centering
\includegraphics[trim={0cm 0cm 0cm 0cm},clip,width=1\columnwidth]{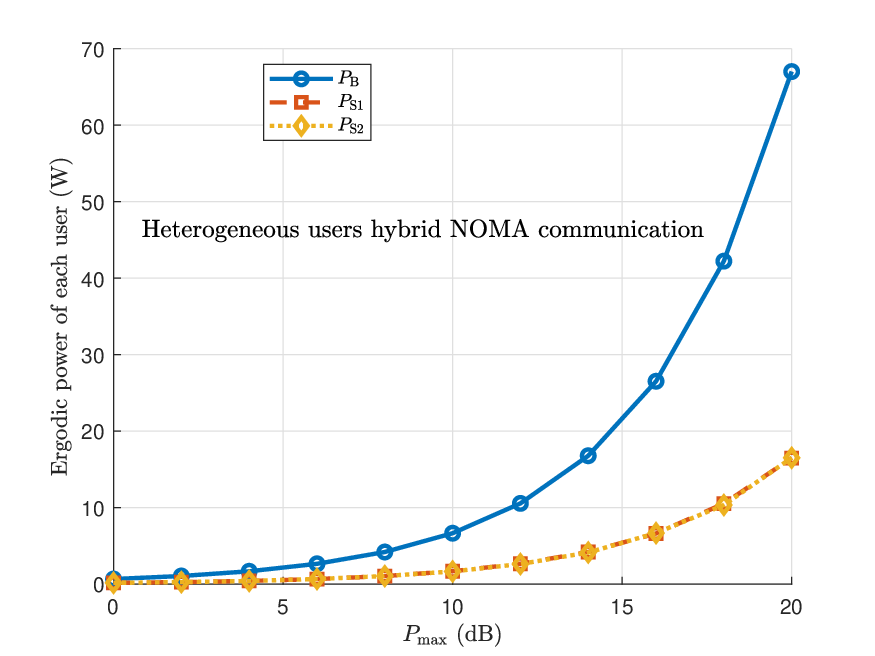}
\caption{Ergodic power of each user versus $P_{\max}$ with $\epsilon_{\mathrm{th}}=0.9$.}
\label{fig10}
\end{figure}

Fig.~\ref{fig10} shows the allocated ergodic power for each user in the hybrid NOMA setup versus $P_{\max}$. The figure shows that with increasing $P_{\max}$ values, the power allocated to the bit user is significantly higher than that for the semantic users. The allocated power for each user increases with $P_{\max}$ but the growth rate for $P_{\text{S1}}$ and $P_{\text{S2}}$ is substantially lower than that for $P_{\text{B}}$. This indicates that the hybrid NOMA system prioritizes bit user which renders optimality to be unfair. This is due to the difference in capacity equations for bit and semantic users, respectively depending on log-rate and a saturating similarity function. 

\begin{figure}[!t]
\centering
\includegraphics[trim={0cm 0cm 0cm 0cm},clip,width=1\columnwidth]{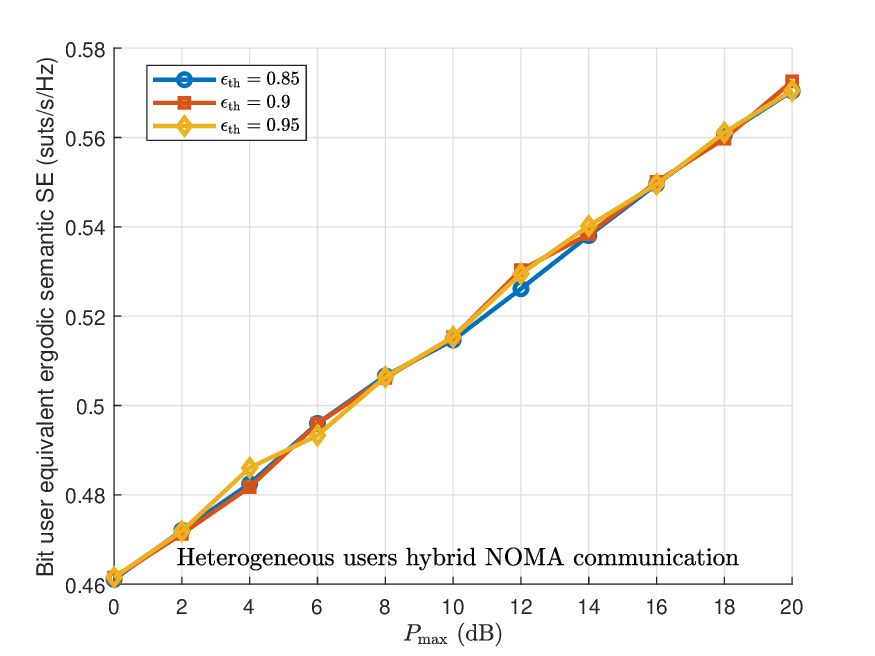}
\caption{Bit user equivalent ergodic semantic SE versus $P_{\max}$ for different $\epsilon_{\mathrm{th}}$.}
\label{fig9a}
\end{figure}

In Fig.~\ref{fig9a}, the equivalent ergodic semantic SE of bit user in the heterogeneous users hybrid NOMA setup is shown versus $P_{\max}$ for $\epsilon_{\mathrm{th}}$ values of 0.85, 0.9, 0.95. The three curves corresponding to different values of $\epsilon_{\mathrm{th}}$ show a steadily growing trend with $P_{\max}$. The figure shows that the equivalent semantic SE of bit user consistently increases with $P_{\max}$. This behavior can be attributed to the efficient resource management of the hybrid NOMA system which prioritizes bit user in a heterogeneous users network for accurate signal reconstruction.

\section{Conclusions}
A novel hybrid NOMA framework is presented to enable the coexistence of semantic and bit users. Higher performance of the proposed heterogeneous users hybrid NOMA setup is demonstrated through analytical evaluation, with simulation results conforming to the related literature on hybrid NOMA and semantic communications. The proposed algorithm dynamically pairs users based on real-time channel conditions and implements a decoding strategy to ensure efficient interference management. The joint power allocation problem was solved under the power constraint to maximize the equivalent ergodic semantic SE. Bit-to-semantic decoding strategy is adopted in slots with both types of users, thereby enabling interference-free decoding of the semantic signal. The simulation results demonstrated the efficacy of the proposed communication setup in enhancing ergodic semantic SE. The proposed system integrates both NOMA and OMA strategies by dynamically assigning users to time slots, ensuring efficient resource utilization and making it a viable approach for heterogeneous users communication in future wireless networks.

\section*{Acknowledgment}
This work was supported by the UK Research and Innovation under the UK government’s
Horizon Europe funding guarantee through MSCA-DN SCION Project Grant
Agreement No.101072375 [grant number: EP/X027201/1].


\end{document}